\documentstyle[epsf]{article}

\font\tenrm=cmr10
\font\tenit=cmti10
\font\elevenbf=cmbx10 scaled\magstep 1
\font\elevenrm=cmr10 scaled\magstep 1
\font\elevenit=cmti10 scaled\magstep 1

\renewenvironment{thebibliography}[1]
 { \elevenrm
   \begin{list}{\arabic{enumi}.}
    {\usecounter{enumi} \setlength{\parsep}{0pt}
     \setlength{\itemsep}{3pt} \settowidth{\labelwidth}{#1.}
     \sloppy
    }}{\end{list}}

\begin{document}
\font\fortssbx=cmssbx10 scaled \magstep2
\hbox to \hsize{
\hskip.5in \raise.1in\hbox{\fortssbx University of Wisconsin - Madison}
\hfill\vbox{\hbox{\bf LBL-35746}
            \hbox{\bf MAD/PH/837}
            \hbox{\bf hep-ph@xxx/9406413}
            \hbox{June 1994}} }

\begin{center}
\vglue 0.6cm
{
 {\elevenbf        \vglue 10pt
         STRONG $WW$ SCATTERING AT PHOTON LINEAR COLLIDERS\footnote{Presented
by MSB at the Workshop on Gamma-gamma Colliders, 28-31 March 1994,
Berkeley, California.
}
\\}
\vglue 1.0cm
{\tenrm M.~S.~Berger$^a$ and M.~S.~Chanowitz$^b$\\}
\baselineskip=13pt
{\tenit $^a$Physics Department, University of Wisconsin\\}
\baselineskip=12pt
{\tenit Madison, WI 53706, USA\\}
\vglue 6pt
{\tenit $^b$Lawrence Berkeley Laboratory\\}
\baselineskip=12pt
{\tenit Berkeley, CA 94720, USA\\}}

\vglue 0.5cm
{\tenrm ABSTRACT}

\end{center}

\vglue 0.3cm
{\rightskip=3pc
 \leftskip=3pc
 \tenrm\baselineskip=12pt
 \noindent
We investigate the possibility of observing strong interactions of
longitudinally polarized weak
vector bosons in the process
$\gamma \gamma \to ZZ$ at a photon linear collider. We make use of polarization
of the photon beams and cuts on the decay products of the $Z$ bosons
to enhance the signal relative to the background of transversely polarized
$ZZ$ pairs. We find that the background overwhelms the signal unless there
are strong resonant effects, as for instance from a technicolor
analogue of the hadronic $f_2(1270)$ meson.}

\vglue 0.4in
\baselineskip=14pt
\elevenrm

\vglue 0.6cm
{\elevenbf\noindent 1. Introduction}
\vglue 0.2cm

Back-scattering of low energy laser photons from the beams of a high energy
$e^+e^-$   linear  collider offers the possibility of obtaining
photon-photon collisions of energy and luminosity comparable to that of the
parent $e^+e^-$ collider.\cite{plc} Here we consider whether such
photon-photon colliders could be used to study a strongly interacting
electroweak symmetry breaking sector in the process $\gamma \gamma \to
Z_L Z_L$, where $Z_L$ denotes a $Z$ boson of longitudinal polarization.
If the symmetry breaking sector is  strongly interacting
$\gamma \gamma \to Z_L Z_L$ is analogous to  the hadronic process
$\gamma \gamma \to \pi^0 \pi^0$.

Other studies of this process have been reported previously.\cite{others}
Here we
focus on whether the  signal can be observed over the background from
transversely polarized $Z$ pairs, using strategies developed for the study
of strong $WW$ scattering at multi-TeV colliders. We conclude that
nonresonant signals are probably not observable but that resonant signals,
such as that of a tensor meson analogous
to the hadron $f_2(1270)$, could be observable with integrated luminosity
of order 100 fb$^{-1}$ and energy near the production threshold. Nonresonant
strong $WW$ dynamics would be better observed at a photon linear collider in
the processes\cite{brodsky-jikia-cheung}
$\gamma \gamma \rightarrow WWWW, ZZWW$,
analogous to the strong scattering process\cite{cg} $qq \rightarrow qqWW$.

The process $\gamma \gamma \to ZZ$  seems more promising than
$\gamma \gamma \to W^+W^-$ which has a large tree-level
background cross section
of order 90 pb, of which even the $W^+_LW^-_L$ is dominated by
the Born approximation near threshold. Since it vanishes in
Born approximation, the process $\gamma \gamma \to ZZ$ is more promising.
However Jikia has shown that the one loop contribution to
$\gamma \gamma \to Z_TZ_T$ is also quite large\cite{jikia}, large enough to
overwhelm the Higgs boson signal in $\gamma \gamma \to H \to ZZ$
for $m_H > 400$ GeV. His calculation has now
been confirmed both analytically\cite{berger} and numerically\cite{ggzz}.
Strong scattering signals are likely to be even smaller, and we find that
the $Z_TZ_T$
background is over two orders of magnitude larger than the typical nonresonant
$\gamma \gamma \to Z_LZ_L$ signal. Drastic improvements must be found
if the signal is to be observable.

\vglue 0.6cm
{\elevenbf\noindent 2. Nonresonant Strong Scattering Signals and Background}
\vglue 0.2cm
The leading contribution to $\gamma \gamma \to ZZ$ occurs via $\gamma \gamma
\to W^+W^-$ with rescattering of the final state into $ZZ$ as for instance in
Fig.~1. This is the dominant contribution both to the nonperturbative
strong interaction $Z_LZ_L$ signal and the perturbative $Z_TZ_T$ background.

In the case of Higgs boson production, $\gamma \gamma \to H \to ZZ$, the
signal cross section decreases with increasing mass $m_H$ while in the
threshold region the background $ZZ$ cross section increases rapidly. At
about $m_H > 400$ GeV the  signal is overwhelmed by the background.
In Higgs boson models the rise of the $W^+_L W^-_L \to Z_L Z_L$ amplitude
(the so-called ``bad high energy behavior'') is cut off by the Higgs boson
exchange amplitude at
$\sqrt{s} = m_H$. In strong scattering models, as considered here, the
cancellation is deferred to the $\sim 2$ TeV scale where the leading $J=0$
partial wave amplitude saturates unitarity. This results in an excess of
$Z_LZ_L$ pairs from the uncancelled gauge sector  $W^+_L W^-_L \to Z_L Z_L$
amplitude that {\elevenit is} the nonresonant strong scattering signal.
The exact
nature of the signal depends on  strong interaction dynamics that
is {\elevenit a
priori} unknown. We will illustrate the nonresonant strong scattering signal
by using two models that when applied to QCD crudely represent, though
underestimate, the experimental
data for $\gamma \gamma \to \pi^0 \pi^0$.

The signal-to-background ratio  improves with increasing energy
 because the $Z_TZ_T$ background becomes more strongly
peaked in the forward and backward directions and so can be more
efficiently removed
by a cut on the scattering angle.  Further gains are possible by using
polarized photon beams chosen to favor the strong scattering signal over
the background. For nonresonant strong scattering the $J=0$ channel
predominates at the energies under consideration and is enhanced by choosing
photon beams of equal helicity, hence  with total $J_Z=0$. The tensor
meson signal is favored by choosing opposite photon helicities that sum
to $J_Z=2$.

As shown previously for strong $WW$ scattering
at hadron colliders\cite{mbmc++},
the longitudinally polarized $Z_LZ_L$ signal can be enhanced over the
transversely polarized  background at large
$\sqrt{s}$ by imposing a $p_T^{}$ cut on the decay products of the $Z$'s.
This discriminates against
transversely  polarized $Z$'s because their decay products
tend to be aligned along the $Z$ boson line-of-flight, while the
decay products of longitudinally polarized $Z$'s are predominantly
perpendicular to the $Z$ line of flight.
(See Figure 2.) The normalized decay distributions are
\begin{equation}
{1\over \Gamma}{{d\Gamma}\over {d\cos \theta ^*}}=
\left \{\begin{array}{l@{\quad}r}
{3\over 4}\sin ^2\theta ^* & \lambda _Z^{}=L \\
{3\over 8}(1+\cos^2\theta ^*) & T
\end{array} \right .
\end{equation}
where $\theta ^*$ is the angle  between the direction in
the $Z$ rest frame
of the decaying quark or lepton and the direction of the $Z$ in the lab.
The decay product moving against the  $Z$ boson line-of-flight will tend to
be soft and will often be rejected by the $p_T$ cut.
The larger $\sqrt{s}$, the  more
highly boosted are the $Z$ bosons and the greater
the effectiveness of the cut.

We consider the decay modes $ZZ\to q\overline{q}\nu \overline{\nu }$,
$\ell^+\ell^-\nu \overline{\nu }$, $ q\overline{q}\ell^+\ell^-$, and
$\ell^+\ell^-\ell^{\prime +}\ell^{\prime -}$,
which account for 40\% of the $Z$
pair decays.
We do not consider the four jet final state because it is likely to be
overwhelmed
by the much larger four jet mode of the $WW$ background, even given excellent
jet-jet mass resolution.
We ignore backgrounds from a $Z$ boson going down the
beam pipe or disappearing into a crack in the detector.
The relevant variables  are the  transverse mass
$M_{ZZ,T}^{}=2\sqrt{M_Z^2+p_{T,Z}^2}$ (where $p_{T,Z}$ is the transverse
momentum of the observed $Z$) and $p_T$, the smallest transverse
momentum of an observable (i.e., not neutrino)
$Z$ decay product.

We use the following three cuts:
(1) $|\cos \theta _{\rm lab}|< \cos \theta_{\rm max}$;
(2) $M_{ZZ,T}>M_{ZZ,T}^{\rm min}$;
(3) $p_T^{}>p_T^{\rm min}$.
For simplicity, in
the $\sim 10$\% of the events in which both Z's can be reconstructed,
$ q\overline{q}\ell^+\ell^-$ and
$\ell^+\ell^-\ell^{\prime +}\ell^{\prime -}$, we arbitrarily designate
one of the $Z$'s
as the ``observable'' one and proceed as above. This reduces the
effectiveness of the cuts
relative to what is actually attainable,
especially  the $p_T$ cut, so that the results
presented below err on the pessimistic side.

To exemplify nonresonant strong scattering we consider two models for
\linebreak $W^+_LW^-_L \to Z_LZ_L$ scattering that
smoothly extrapolate the threshold behavior required by the low energy
theorems\cite{cg,cgg} in a manner consistent with unitarity.
The linear model extrapolates the
{\elevenit absolute value} of the $J=0$ partial wave amplitude. Applied to
$\pi^+ \pi^- \to \pi^0 \pi^0$ scattering it is
\begin{eqnarray}
|a_0^{}|&=& X \cdot \theta\left ({2\over 3}\ -\ X\right )\ +\ {2 \over 3}
\cdot \theta\left (X - {2 \over 3} \right )\;,
\end{eqnarray}
where $X = (s-m_\pi^2)/16 \pi F_\pi^2$. The cross section is then\cite{dhl,dp}
\begin{eqnarray}
\sigma &=&{{\beta \alpha ^2}\over {\pi s}}
|a_{0}^{}|^2\;.
\end{eqnarray}
The factor 2/3 follows from decomposing $a_0(\pi^+ \pi^- \to \pi^0 \pi^0)$
into isospin channels $a_{IJ}^{}$, that is,
$a_0 = {2\over 3}(a_{00}^{}-a_{20}^{})$,
and applying elastic unitarity to the $a_{IJ}^{}$.

The second model uses the more gradual K-matrix unitarization,
\begin{eqnarray}
a_{00}^{}&=&X_0 (1-iX_0 )^{-1}\;,
\\
a_{20}^{}&=&-X_2 (1+iX_2 )^{-1}\;,
\end{eqnarray}
where $X_0 = (s - {m_{\pi}^2 \over 2}) / 16 \pi F_\pi^2$ and
$X_2 = (s - 2m_{\pi}^2) / 32 \pi F_\pi^2$.
The cross section for $\gamma \gamma \to \pi ^0\pi ^0$ is
\begin{eqnarray}
\sigma &=&{{\beta \alpha ^2}\over {\pi s}}\left ({2\over 3}
|a_{00}^{}-a_{20}^{}|\right )^2\;.
\end{eqnarray}

The models slightly underestimate the experimental data
from the Crystal Ball\cite{crystalball} below 700 MeV
as shown in Figure 3. We apply the
models to \linebreak $W_L^+W_L^- \to Z_LZ_L$ by setting
$m_\pi=0$ and taking $F_\pi \to v$.
The conclusions given below would not change even if the amplitudes were
increased by a factor of two.

We convolute the gamma-gamma
cross sections
with the luminosity
distributions to obtain the total cross sections given in
Tables 1 and 2, which do not include decay branching ratios.
We consider two cases:
(i) unpolarized beams and (ii) polarized photon and electron beams,
$\lambda _{\gamma}=-1$ and $\lambda _e=1/2$.
In Tables 1 and 2  the strong scattering signal is
less than $1$ fb.
Since the $TT$ background is $260$ fb, the signal is
more than two orders of magnitude smaller than the background before any cuts
are made.

The correlations between the final state fermions are obtained by using the
decay density matrix $\rho $ for each $Z$ boson,
\begin{eqnarray}
\sigma_{\lambda _1\lambda _2}&=&\sum _{\lambda _3\lambda _3^{\prime}
\lambda _4\lambda _4^{\prime}}\int
{\cal M}_{\lambda _1\lambda _2\lambda _3\lambda _4}
{\cal M}_{\lambda _1\lambda _2\lambda _3^{\prime}\lambda _4^{\prime}}^*
\rho_{\lambda _3\lambda _3^{\prime}}\rho_{\lambda _4\lambda _4^{\prime}}
{\rm dLIPS}\;.
\end{eqnarray}
The results are summarized in tables 3 and 4 where we exhibit the number
of signal and background events for 100 fb$^{-1}$ with fixed angle cut
$\cos(\theta_{\rm lab}) < 0.7$ and with the cuts on $p_T$ and $M_{ZZ,T}$
chosen to optimize the statistical significance of the signal.
We have investigated increasing the angle cut to $|\cos \theta _{\rm lab}|<0.6$
and find that this does not increase the statistical significance of the
signal. The $S:B$ ratios are too small to be observable, even if the
statistical significance $S/\sqrt{B}$ were enhanced by imagining still
higher luminosity.

\vglue 0.6cm
{\elevenbf\noindent 3. Tensor Resonance}
\vglue 0.2cm

Far enough above threshold $J \neq 0$ partial waves begin to contribute to
$\gamma \gamma \to Z_L Z_L$ scattering. Just as the d-wave begins to emerge in
$\pi \pi$ scattering above 1 GeV, we might expect d-wave scattering in
$\gamma \gamma \to Z_L Z_L$ to set in above about $v/F_{\pi}
\cdot 1 {\rm GeV} \sim 2.5$ TeV. In hadron physics the d-wave
saturates unitarity
at the peak of the $f_2(1270)$ resonance, which in
$SU(3)_{TC}$ would correspond
to $\sim 3.4$ TeV. Figure 4 shows that the experimental cross section for
$\gamma \gamma \to \pi^0 \pi^0$ is nearly two orders of magnitude
larger than the nonresonant
models at the peak of the $f_2$ and a few times larger
in the region of the $f_0(975)$ scalar resonance.

To illustrate a resonant signal we use the experimental
data for $\gamma \gamma \to f_2(1270) \to \pi^0 \pi^0$ with the energy
rescaled by $v/F_{\pi} \sim 2700$, as expected in a technicolor
theory with an $SU(3)$ gauge group. Using conventional large $N$ scaling
estimates we also consider the $SU(5)_{\rm TC}$ technicolor gauge group which
implies lower mass, and therefore more easily observable, techni-hadrons.
The mass and width of the technitensor are then obtained from the hadron
$f_2(1270)$,
\begin{eqnarray}
M_{f_{TC}^{}}&=&\sqrt{3\over {N_{TC}}}{v\over {f_{\pi}}}M_f\;, \\
\Gamma _{f_{TC}^{}}&=&{3\over {N_{TC}}}{{M_{f_{TC}^{}}}\over {M_f}}\Gamma _f\;.
\end{eqnarray}

We assume the peak at 1270 MeV in the data
is comprised of the ratios $D_2$:$D_0$:$S$=0.8:0.2:0. That is
we assume for simplicity that the peak has only a tensor component,
despite the fact that there is evidence for a $J=0$ component.
Since the helicity of the QCD tensor is predominantly $J_Z^{}=2$,
unpolarized beams are preferable to the polarized beams used for the
s-wave contributions. Polarization settings that isolate the $J_Z^{}=2$
helicity, may improve the situation beyond that obtained with unpolarized
beams.

The signal (for $N_{TC} = 3$) and background are displayed in
Fig.~5 for  unpolarized
beams at a $\sqrt{s_{e^+e^-}}=4$ TeV collider (which has maximum
$\gamma \gamma$ energy $\sqrt{s_{\gamma \gamma}} \sim 3.2$ TeV).
The signal is smaller
than it would  be for $N_{TC}^{}>3$ because the  peak is
partially beyond the reach of the $\gamma \gamma$ center-of-mass energy.
The statistical significance of the signal is maximized with a transverse
mass cut $M_{ZZ,T} > 2800$ GeV.
With 100 fb$^{-1}$ of integrated luminosity this yields 28.6 signal events
and 34.8 background events.

For $N_{TC}^{}=5$ the situation is improved considerably. The tensor peak is
fully within reach of the 4 TeV machine,
and the peak is more pronounced due to the
$(N_{TC}/3)^{3\over 2}$ enhancement. See Figure 6. We find that the
statistical significance of the signal is maximized with the
cuts $M_{ZZ,T} > 2800$ GeV and $p_T^{}>180$ GeV, for which there are 221.9
signal events and 240.8 background events. The tensor peak for the case
of $SU(5)_{TC}$ lies at $\sim 2.6$ TeV, so is partially visible at a 3 TeV
$\sqrt{s_{e^+e^-}}$ collider, for which we find 46.3 signal events and 52.0
background events (relaxing the $p_T^{}$ cut slightly to 170 GeV).

\vglue 0.6cm
{\elevenbf\noindent 4. Conclusions}
\vglue 0.2cm
In summary our conclusions are
\begin{itemize}
\item $S/B$ for the nonresonant s-wave is small,
even for $\sqrt{s_{e^+e^-}}=4$ TeV.

\item Significant signals are possible if there is a tensor resonance
within the energy reach of the collider.

\item The effect of  fermionic cuts on the heavy
Higgs signal should be investigated. We expect that the utility of this
method is limited since the signal is falling rapidly and the background is
growing rapidly at the limit of feasibility, $M_H^{}\sim 400 $ GeV.
The $Z$ bosons are less highly boosted so that the fermionic cuts are
less efficient.

\item Nonresonant strong scattering
might  be  better observed in  $\gamma \gamma \to WWWW$ and
$\gamma \gamma \to WWZZ$\cite{brodsky-jikia-cheung}.

\end{itemize}

\vglue 0.6cm
{\elevenbf\noindent 5. Acknowledgements}
\vglue 0.2cm
MSC's work was supported by the Director, Office of Energy Research, Office of
High Energy and Nuclear Physics, Division of High Energy Physics of the U.S.
Department of Energy under Contract DE-AC03-76SF0.
MSB's work was supported
in part by the University of Wisconsin Research Committee with funds granted by
the Wisconsin Alumni Research Foundation, in part by the U.S.~Department of
Energy under contract no.~DE-AC02-76ER00881, and in part by the Texas National
Laboratory Research Commission under grant no.~RGFY93-221 and
by an SSC Fellowship.

\vglue 0.6cm
{\elevenbf\noindent 6. References}
\vglue 0.2cm

\newpage

{\center \begin{tabular}{|c|c|c|}
\hline
\multicolumn{1}{|c|}{$\sqrt{s_{e^+e^-}}$}
&\multicolumn{1}{|c|}{Polarization}
&\multicolumn{1}{|c|}{Cross Section [fb]}
\\ \hline \hline
\multicolumn{1}{|c|}{1.5 TeV}
&\multicolumn{1}{|c|}{Polarized}
&\multicolumn{1}{|c|}{0.54}
\\ \cline{2-3}
\multicolumn{1}{|c|}{}
&\multicolumn{1}{|c|}{Nonpolarized}
&\multicolumn{1}{|c|}{0.34}
\\ \hline \hline
\multicolumn{1}{|c|}{2 TeV}
&\multicolumn{1}{|c|}{Polarized}
&\multicolumn{1}{|c|}{0.70}
\\ \cline{2-3}
\multicolumn{1}{|c|}{}
&\multicolumn{1}{|c|}{Nonpolarized}
&\multicolumn{1}{|c|}{0.52}
\\ \hline \hline
\multicolumn{1}{|c|}{3 TeV}
&\multicolumn{1}{|c|}{Polarized}
&\multicolumn{1}{|c|}{0.56}
\\ \cline{2-3}
\multicolumn{1}{|c|}{}
&\multicolumn{1}{|c|}{Nonpolarized}
&\multicolumn{1}{|c|}{0.65}
\\ \hline \hline
\multicolumn{1}{|c|}{4 TeV}
&\multicolumn{1}{|c|}{Polarized}
&\multicolumn{1}{|c|}{0.36}
\\ \cline{2-3}
\multicolumn{1}{|c|}{}
&\multicolumn{1}{|c|}{Nonpolarized}
&\multicolumn{1}{|c|}{0.59}
\\ \hline
\end{tabular}
\vskip .3in }
\begin{center}
{\bf Table 1:}  K Matrix Cross Sections without any cuts.
\end{center}
\vspace*{2in}
{\center \begin{tabular}{|c|c|c|}
\hline
\multicolumn{1}{|c|}{$\sqrt{s_{e^+e^-}}$}
&\multicolumn{1}{|c|}{Polarization}
&\multicolumn{1}{|c|}{Cross Section [fb]}
\\ \hline \hline
\multicolumn{1}{|c|}{1.5 TeV}
&\multicolumn{1}{|c|}{Polarized}
&\multicolumn{1}{|c|}{0.65}
\\ \cline{2-3}
\multicolumn{1}{|c|}{}
&\multicolumn{1}{|c|}{Nonpolarized}
&\multicolumn{1}{|c|}{0.38}
\\ \hline \hline
\multicolumn{1}{|c|}{2 TeV}
&\multicolumn{1}{|c|}{Polarized}
&\multicolumn{1}{|c|}{0.94}
\\ \cline{2-3}
\multicolumn{1}{|c|}{}
&\multicolumn{1}{|c|}{Nonpolarized}
&\multicolumn{1}{|c|}{0.65}
\\ \hline \hline
\multicolumn{1}{|c|}{3 TeV}
&\multicolumn{1}{|c|}{Polarized}
&\multicolumn{1}{|c|}{0.60}
\\ \cline{2-3}
\multicolumn{1}{|c|}{}
&\multicolumn{1}{|c|}{Nonpolarized}
&\multicolumn{1}{|c|}{0.77}
\\ \hline \hline
\multicolumn{1}{|c|}{4 TeV}
&\multicolumn{1}{|c|}{Polarized}
&\multicolumn{1}{|c|}{0.47}
\\ \cline{2-3}
\multicolumn{1}{|c|}{}
&\multicolumn{1}{|c|}{Nonpolarized}
&\multicolumn{1}{|c|}{0.72}
\\ \hline
\end{tabular}
\vskip .3in }
\begin{center}
{\bf Table 2:}  Linear Model Cross Sections without any cuts.
\end{center}
\vskip .5in
\newpage
{\center \begin{tabular}{|c|c|c|c|c|c|c|}
\hline
\multicolumn{1}{|c|}{$\sqrt{s_{e^+e^-}}$}
&\multicolumn{1}{|c|}{Model}
&\multicolumn{1}{|c|}{$M_{ZZT}$ Cut}
&\multicolumn{1}{|c|}{$p_T$ Cut}
&\multicolumn{1}{|c|}{Signal $S$}
&\multicolumn{1}{|c|}{Background $B$}
&\multicolumn{1}{|c|}{Stat. Sig. $S/\sqrt{B}$}
\\ \hline \hline
\multicolumn{1}{|c|}{2 TeV}
&\multicolumn{1}{|c|}{Linear}
&\multicolumn{1}{|c|}{850}
&\multicolumn{1}{|c|}{110}
&\multicolumn{1}{|c|}{10.1}
&\multicolumn{1}{|c|}{708.8}
&\multicolumn{1}{|c|}{0.38}
\\ \cline{2-7}
\multicolumn{1}{|c|}{}
&\multicolumn{1}{|c|}{K Unit.}
&\multicolumn{1}{|c|}{800}
&\multicolumn{1}{|c|}{90}
&\multicolumn{1}{|c|}{8.3}
&\multicolumn{1}{|c|}{866.6}
&\multicolumn{1}{|c|}{0.28}
\\ \hline
\multicolumn{1}{|c|}{3 TeV}
&\multicolumn{1}{|c|}{Linear}
&\multicolumn{1}{|c|}{900}
&\multicolumn{1}{|c|}{110}
&\multicolumn{1}{|c|}{11.9}
&\multicolumn{1}{|c|}{781.9}
&\multicolumn{1}{|c|}{0.42}
\\ \cline{2-7}
\multicolumn{1}{|c|}{}
&\multicolumn{1}{|c|}{K Unit.}
&\multicolumn{1}{|c|}{900}
&\multicolumn{1}{|c|}{110}
&\multicolumn{1}{|c|}{9.9}
&\multicolumn{1}{|c|}{781.9}
&\multicolumn{1}{|c|}{0.36}
\\ \hline
\multicolumn{1}{|c|}{4 TeV}
&\multicolumn{1}{|c|}{Linear}
&\multicolumn{1}{|c|}{900}
&\multicolumn{1}{|c|}{130}
&\multicolumn{1}{|c|}{9.4}
&\multicolumn{1}{|c|}{661.0}
&\multicolumn{1}{|c|}{0.36}
\\ \cline{2-7}
\multicolumn{1}{|c|}{}
&\multicolumn{1}{|c|}{K Unit.}
&\multicolumn{1}{|c|}{900}
&\multicolumn{1}{|c|}{130}
&\multicolumn{1}{|c|}{7.8}
&\multicolumn{1}{|c|}{661.0}
&\multicolumn{1}{|c|}{0.30}
\\ \hline
\end{tabular}
\vskip .3in }
\begin{center}
{\bf Table 3:}  Event rates at unpolarized colliders with
angle cut $\cos(\theta_{\rm lab})<0.7$.
\end{center}
\vskip .5in

\vspace*{2in}

{\center \begin{tabular}{|c|c|c|c|c|c|c|}
\hline
\multicolumn{1}{|c|}{$\sqrt{s_{e^+e^-}}$}
&\multicolumn{1}{|c|}{Model}
&\multicolumn{1}{|c|}{$M_{ZZT}$ Cut}
&\multicolumn{1}{|c|}{$p_T$ Cut}
&\multicolumn{1}{|c|}{Signal $S$}
&\multicolumn{1}{|c|}{Background $B$}
&\multicolumn{1}{|c|}{Stat. Sig. $S/\sqrt{B}$}
\\ \hline \hline
\multicolumn{1}{|c|}{2 TeV}
&\multicolumn{1}{|c|}{Linear}
&\multicolumn{1}{|c|}{1050}
&\multicolumn{1}{|c|}{120}
&\multicolumn{1}{|c|}{18.8}
&\multicolumn{1}{|c|}{907.4}
&\multicolumn{1}{|c|}{0.62}
\\ \cline{2-7}
\multicolumn{1}{|c|}{}
&\multicolumn{1}{|c|}{K Unit.}
&\multicolumn{1}{|c|}{1050}
&\multicolumn{1}{|c|}{110}
&\multicolumn{1}{|c|}{14.0}
&\multicolumn{1}{|c|}{961.8}
&\multicolumn{1}{|c|}{0.45}
\\ \hline
\multicolumn{1}{|c|}{3 TeV}
&\multicolumn{1}{|c|}{Linear}
&\multicolumn{1}{|c|}{750}
&\multicolumn{1}{|c|}{140}
&\multicolumn{1}{|c|}{11.9}
&\multicolumn{1}{|c|}{823.4}
&\multicolumn{1}{|c|}{0.41}
\\ \cline{2-7}
\multicolumn{1}{|c|}{}
&\multicolumn{1}{|c|}{K Unit.}
&\multicolumn{1}{|c|}{750}
&\multicolumn{1}{|c|}{140}
&\multicolumn{1}{|c|}{11.1}
&\multicolumn{1}{|c|}{823.4}
&\multicolumn{1}{|c|}{0.38}
\\ \hline
\multicolumn{1}{|c|}{4 TeV}
&\multicolumn{1}{|c|}{Linear}
&\multicolumn{1}{|c|}{750}
&\multicolumn{1}{|c|}{120}
&\multicolumn{1}{|c|}{9.0}
&\multicolumn{1}{|c|}{773.1}
&\multicolumn{1}{|c|}{0.32}
\\ \cline{2-7}
\multicolumn{1}{|c|}{}
&\multicolumn{1}{|c|}{K Unit.}
&\multicolumn{1}{|c|}{650}
&\multicolumn{1}{|c|}{110}
&\multicolumn{1}{|c|}{6.6}
&\multicolumn{1}{|c|}{817.4}
&\multicolumn{1}{|c|}{0.23}
\\ \hline
\end{tabular}
\vskip .3in }
\begin{center}
{\bf Table 4:}  Event rates at polarized colliders with
angle cut $\cos(\theta_{\rm lab})<0.7$.
\end{center}
\vskip .5in

\newpage

\begin{center}
\epsfxsize=3in
\hspace{0in}
\epsffile{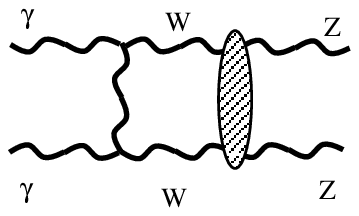}

\parbox{5.5in}{\small \small Fig.~1. The process $\gamma \gamma \to Z_LZ_L$
occurs via $\gamma \gamma \to W^+_LW^-_L$ with a final state interaction by
strong $W_LW_L$ scattering, as shown for example in the Feynman diagram
above.}
\end{center}

\vspace*{2in}
\begin{center}
\epsfxsize=3in
\hspace{0in}
\epsffile{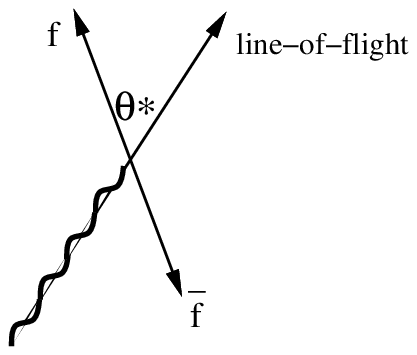}

\parbox{5.5in}{\small \small Fig.~2. The transversely polarized $Z$ bosons tend
to decay with one product along and one against the line-of-flight.}
\end{center}

\newpage
\begin{center}
\epsfxsize=6in
\hspace{0in}
\epsffile{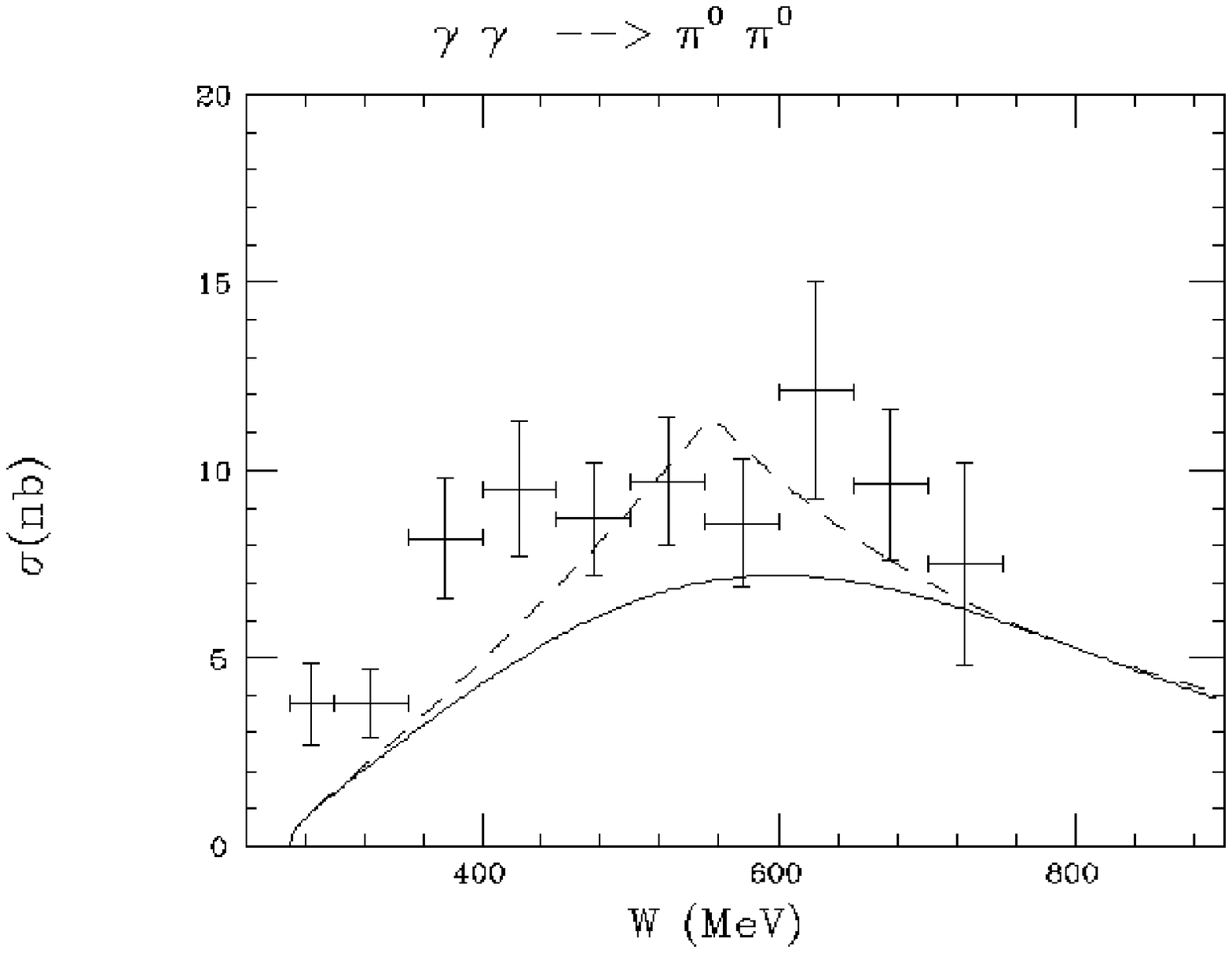}

\parbox{5.5in}{\small \small Fig.~3. The Crystal Ball data\cite{crystalball}
with the linear (dashed) and K-matrix (solid) unitarized models for
the $\gamma \gamma \to \pi^0\pi^0$ process
($W\equiv \sqrt{s}$).}
\end{center}

\begin{center}
\epsfxsize=6in
\hspace{0in}
\epsffile{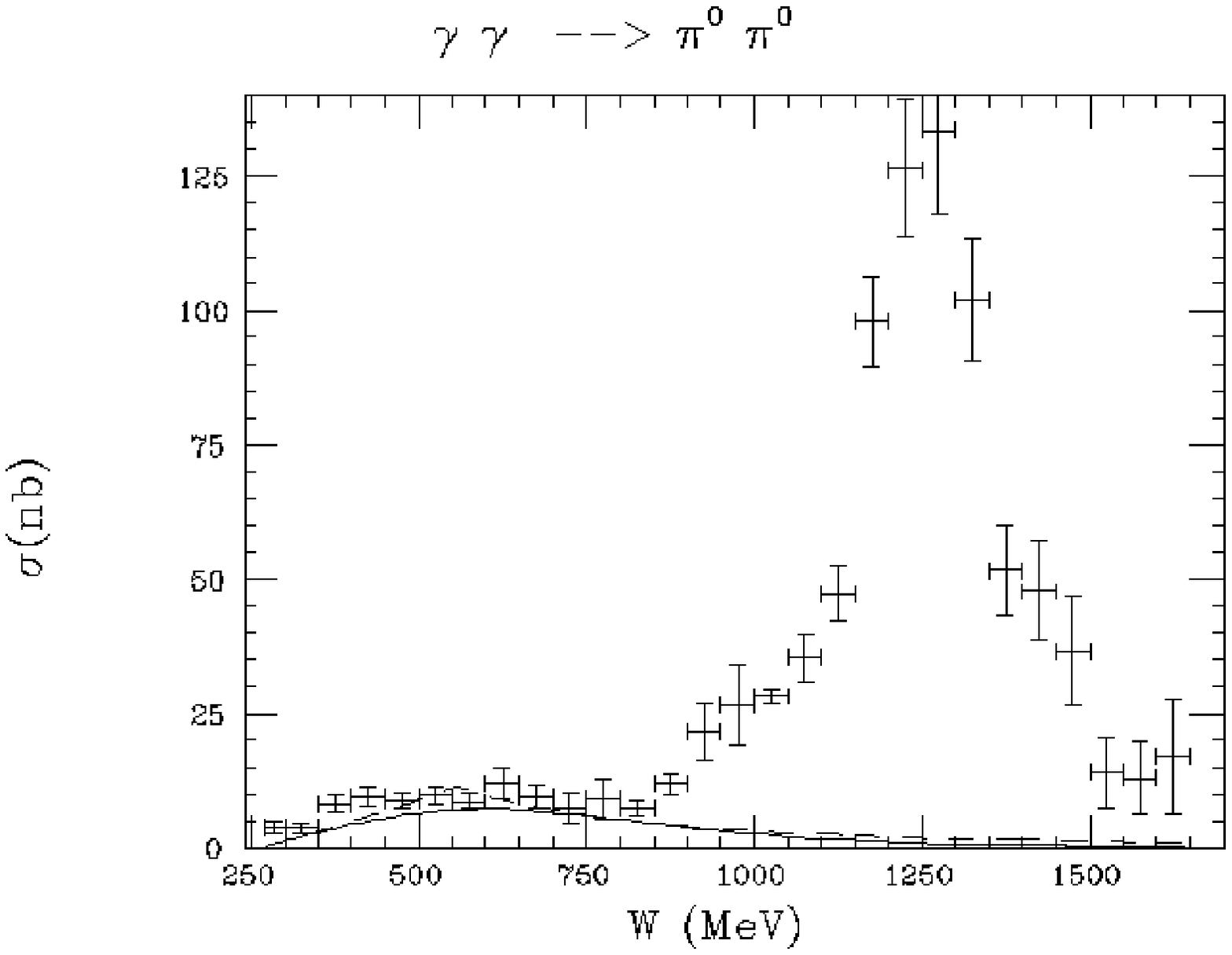}

\parbox{5.5in}{\small \small Fig.~4. The Crystal Ball data with the linear
and K-matrix unitarized models for the $\gamma \gamma \to \pi^0\pi^0$ process.
The tensor $f_2(1270)$ and the scalar $f_0(975)$ are visible
($W\equiv \sqrt{s}$).}
\end{center}

\begin{center}
\epsfxsize=5in
\hspace{0in}
\epsffile{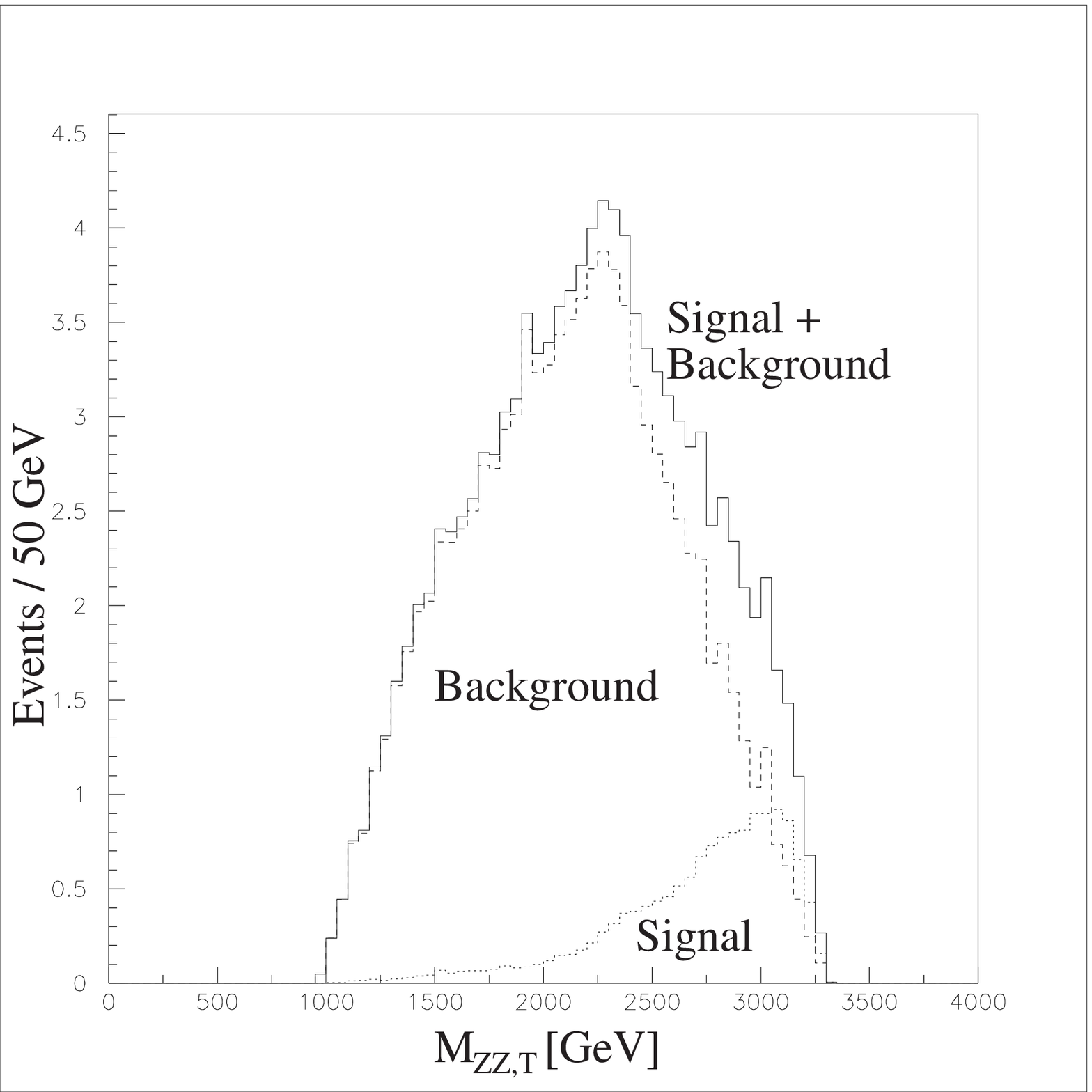}
\vspace{0.2in}

\parbox{5.5in}{\small \small Fig.~5. The tensor contribution produces an
enhancement over the purely s-wave contributions from the low-energy theorem.
The signal and background are shown after the cuts
$|\cos \theta _{\rm lab}|<0.7$ and $p_T^{}>240$ GeV for 10 fb$^{-1}$ of
integrated luminosity without including the branching ratios.}
\end{center}

\newpage
\begin{center}
\epsfxsize=5in
\hspace{0in}
\epsffile{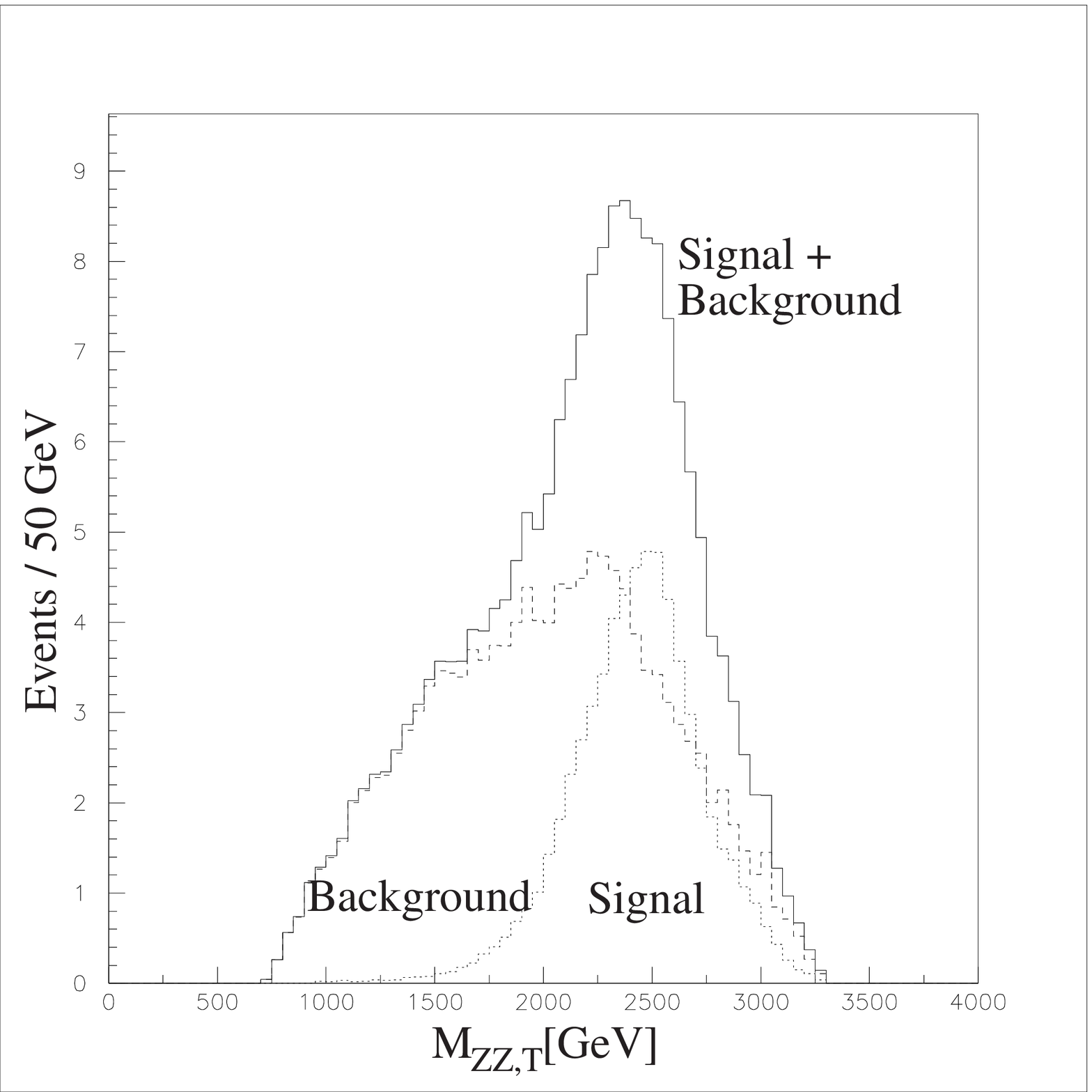}
\vspace{0.2in}

\parbox{5.5in}{\small \small Fig.~6. The tensor contribution for $N_{TC}=5$.
The signal and background are shown after the cuts
$|\cos \theta _{\rm lab}|<0.7$ and $p_T^{}>180$ GeV for 10 fb$^{-1}$ of
integrated luminosity without including the branching ratios.}
\end{center}

\end{document}